# Agency-Driven Labor Theory: A Framework for Understanding Human Work in the AI Age

Venkat Ram Reddy Ganuthula
Indian Institute of Technology Jodhpur

**Abstract**

This paper introduces Agency-Driven Labor Theory (ADLT) as a new theoretical framework for understanding human work in AI-augmented environments. While traditional labor theories have focused primarily on task execution and labor time, ADLT proposes that human labor value is increasingly derived from agency—the capacity to make informed judgments, provide strategic direction, and design operational frameworks for AI systems. The paper presents a mathematical framework expressing labor value as a function of agency quality, direction effectiveness, and outcomes [LV = f(A, D, O)], providing a quantifiable approach to analyzing human value creation in AI-augmented workplaces. Drawing on recent work in organizational economics and knowledge worker productivity, ADLT explains how human workers create value by orchestrating complex systems that combine human and artificial intelligence. The theory has significant implications for job design, compensation structures, professional development, and labor market dynamics. Through applications across various sectors, the paper demonstrates how ADLT can guide organizations in managing the transition to AI-augmented operations while maximizing human value creation. The framework provides practical tools for policymakers and educational institutions as they prepare workers for a labor market where value creation increasingly centers on agency and direction rather than execution.

## I. Introduction

The rapid advancement of artificial intelligence technologies is fundamentally transforming the nature of human work in unprecedented ways. Unlike previous technological revolutions that primarily automated routine manual tasks (Autor et al., 2020), AI systems are increasingly capable of performing complex cognitive tasks that were once the exclusive domain of human knowledge workers. This transformation challenges our traditional understanding of labor value creation and necessitates a fundamental reconceptualization of how we theorize about human work in the modern economy.

Traditional labor theories, from Smith's division of labor to Marx's labor theory of value and more recent human capital frameworks, have primarily focused on direct execution of tasks as the core driver of value creation (Becker, 2009). While these theories have served us well in understanding industrial-era labor dynamics, they prove increasingly inadequate in capturing the evolving nature of human work in an AI-augmented environment. The key limitation of existing frameworks lies in their emphasis on labor time and task execution rather than the uniquely human capabilities of agency, judgment, and strategic direction that are becoming increasingly central to value creation.

This paper introduces Agency-Driven Labor Theory (ADLT) as a new theoretical framework for understanding human work in the AI age. At its core, ADLT proposes that the primary source of human labor value is shifting from direct task execution to the exercise of agency – the capacity to make informed judgments, provide strategic direction, and design frameworks for AI systems to operate within. This shift represents a fundamental evolution in how human workers create value, moving from being primary executors to becoming orchestrators of complex systems that combine human and artificial intelligence.

The mathematical foundations of ADLT provide a formal structure for understanding this transformation. By expressing labor value as a function of agency quality, direction effectiveness, and outcomes [LV = f(A, D, O)], we create a quantifiable framework for analyzing how human workers create value in AI-augmented environments. This approach builds upon recent work in organizational economics (Roberts, 2018) and knowledge worker productivity (Davenport, 2021) while introducing novel constructs for measuring and evaluating human contribution in increasingly automated workplaces.

The implications of this theoretical framework extend beyond academic discourse. As organizations grapple with questions of job design, compensation structures, and skill development in an AI-augmented workplace, ADLT provides practical tools for understanding and managing these transitions. Furthermore, the framework offers insights for policymakers and educational institutions as they work to prepare current and future generations for a labor market where human value creation increasingly centers on agency and direction rather than execution.

This paper proceeds as follows: Section II establishes the theoretical foundations of ADLT, drawing on existing literature in labor economics, organizational theory, and knowledge work. Section III presents the core mathematical framework, including detailed formulations of agency quality and direction effectiveness. Section IV demonstrates empirical applications and measurement approaches, while Sections V through VIII explore organizational implications, future of work considerations, policy recommendations, and theoretical extensions. We conclude with a discussion of key contributions and directions for future research.

## II. Theoretical Foundations

### A. Historical Context

The evolution of labor theory reflects humanity's ongoing attempt to understand and quantify the value creation process in work. Classical economic thought, beginning with Adam Smith's (1776) analysis of the division of labor, established the foundational principle that human work could be systematically analyzed and optimized. This early framework emphasized specialization and skill development as key drivers of productivity, a perspective that would prove particularly relevant to the industrial revolution but requires significant revision in the AI age.

The subsequent development of Marx's labor theory of value marked a crucial theoretical advancement by explicitly connecting human effort to economic value creation. While Marx's specific formulations have faced substantial criticism (Robinson, 1962; Samuelson, 1971), his core insight about the relationship between human activity and value production remains relevant to understanding modern knowledge work. However, Marx's focus on labor time as the primary measure of value proves increasingly inadequate in an era where value creation often depends more on judgment quality than time invested.

The neoclassical revolution in economics brought marginal productivity theory, which provided a more nuanced framework for understanding labor value in market economies. Scholars like Marshall (1890) and later Solow (1956) developed sophisticated models of how labor contributes to production, but their frameworks primarily focused on labor as a relatively homogeneous input into production functions. This approach, while mathematically elegant, struggles to capture the complexity of human agency in modern knowledge work.

The human capital revolution, pioneered by Schultz (1961) and Becker (1964), represented a significant theoretical advancement by recognizing that labor value depends heavily on accumulated knowledge and skills. This framework proved particularly valuable for understanding knowledge work in the late 20th century, but its emphasis on traditional skill acquisition and educational credentials may not fully capture the evolving nature of human value creation in AI-augmented workplaces.

More recent theoretical developments in knowledge work theory (Drucker, 1999; Davenport, 2005) have attempted to address the unique characteristics of cognitive labor, emphasizing the importance of decision-making and creativity. However, these frameworks were developed primarily for an era where humans were the primary performers of complex cognitive tasks. The emergence of AI systems capable of performing many traditional knowledge work tasks creates significant theoretical gaps in our understanding of human labor value.

These gaps in existing theory become particularly apparent when we consider three key aspects of modern work:

1. The increasing importance of framework design and system orchestration over direct task execution
2. The evolving role of human judgment in AI-augmented decision processes
3. The growing value of exception handling and novel situation management

Traditional labor theories provide limited guidance for understanding these aspects of modern work, as they generally assume human workers as primary task executors rather than system orchestrators. This theoretical limitation has practical implications for organization design, compensation systems, and professional development frameworks.

The emergence of AI as a powerful complement to human labor thus creates an urgent need for new theoretical frameworks. While existing theories provide valuable insights into aspects of human work, they fail to fully capture the evolving nature of value creation in AI-augmented environments. Agency-Driven Labor Theory (ADLT) builds upon these historical foundations while introducing new constructs specifically designed to understand human work in the AI age.

**III. Mathematical Framework of ADLT**

**A. Core Mathematical Formulation**

The mathematical framework of Agency-Driven Labor Theory (ADLT) provides a formal structure for analyzing how human workers create value in AI-augmented environments. At its core, ADLT expresses labor value through a primary function that incorporates three fundamental components: agency quality, direction effectiveness, and outcomes.

The primary labor value function is expressed as:

$$LV = f(A, D, O)$$

Where:

- $LV$ represents Labor Value
- $A$ represents Agency Quality
- $D$ represents Direction Effectiveness
- $O$ represents Outcomes

This foundational equation builds upon traditional labor value theories but introduces crucial new elements that capture the evolving nature of human work. Each component is further decomposed into measurable subcomponents that allow for practical application and empirical testing.

The Agency Quality component ($A$) is expressed through a weighted function of three key elements:

$$A = \alpha(F_s + E_d + Q_j)$$

Where:

- Fs represents the Framework design score
- Ed represents the Exception detection rate
- Qj represents the Quality of judgment
- α represents the Agency effectiveness coefficient

This formulation captures the essential elements of human agency in AI-augmented environments. The Framework design score (Fs) measures the worker's ability to create effective operational frameworks for AI systems. The Exception detection rate (Ed) quantifies the capacity to identify situations requiring human intervention. The Quality of judgment (Qj) evaluates the effectiveness of human decision-making in non-standard situations.

**B. Core Principles with Mathematical Expression**

Building upon the primary formulation, ADLT develops several key principles that describe how labor value evolves in AI-augmented environments. These principles are expressed through detailed mathematical relationships that capture the dynamic nature of human-AI interaction.

1. Labor Value Proposition

The detailed form of Agency Quality is expressed as:

$$AQ = \sum(w_i \times a_i)$$

Where:

- $w_i$ represents the weight assigned to each agency component
- $a_i$ represents individual agency metrics

Direction Effectiveness is similarly quantified:

$$DE = \sum(v_i \times d_i)$$

Where:

- $v_i$ represents the weight of each direction component
- $d_i$ represents individual direction metrics

Outcomes are expressed through a multiplicative function:

$$O = \gamma(R_v \times S_v \times I_n)$$

Where:

- $R_v$ represents realized value

- Sv represents scalability value
- In represents innovation contribution
- γ represents the outcome effectiveness coefficient

2. Value Creation Mechanisms

The framework incorporates several key mechanisms that describe how value accumulates over time:

Framework Multiplication Effect: $FM = Fs \times (1 + Ad)^t$

Where:

- Ad represents the adoption rate of the framework
- t represents time
- Fs represents the initial framework score

This equation captures how well-designed frameworks can create exponential value through widespread adoption and iteration.

Exception Handling Premium: $EH = \beta(Ed \times resolution\_rate)$

Where:

- β represents the premium coefficient
- Ed represents the exception detection rate
- resolution_rate represents the successful handling of exceptions

This formulation quantifies the additional value created through effective management of non-standard situations.

Agency Growth: $dA/dt = k(current\_agency - baseline)$

This differential equation describes how agency quality evolves over time, where:

- k represents the learning rate coefficient
- baseline represents the minimum agency level for the role

These mathematical relationships provide a structured framework for analyzing and measuring human value creation in AI-augmented work environments. The framework is designed to be both theoretically rigorous and practically applicable, allowing for empirical testing and refinement through real-world application.

## IV. Empirical Applications and Measurement

### A. Role Transformation Analysis with Metrics

The practical application of ADLT requires systematic measurement approaches across different professional domains. We present detailed metrics frameworks for two broad categories of knowledge work, demonstrating how the theoretical constructs can be operationalized in practice.

1. Knowledge Work Analysis

For knowledge work roles, the value creation function takes the form:

$$KW = A(k) \times D(k) \times O(k)$$

Where $A(k)$, $D(k)$, and $O(k)$ represent domain-specific implementations of agency, direction, and outcomes respectively. We examine three representative domains:

Research Analysis:

- Agency metrics: Novel insight generation rate, methodology innovation score
- Direction metrics: Research framework effectiveness, team alignment index
- Outcome metrics: Citation impact, practical application rate
- Composite score: RA = (0.4 × agency_score + 0.3 × direction_score + 0.3 × outcome_score)

Software Engineering:

- Agency metrics: Architecture design quality, technical debt management
- Direction metrics: Team productivity multiplier, code reuse effectiveness
- Outcome metrics: System reliability, scalability index
- Composite score: SE = (0.35 × agency_score + 0.35 × direction_score + 0.3 × outcome_score)

Financial Management:

- Agency metrics: Strategy formation quality, risk assessment accuracy
- Direction metrics: Portfolio optimization effectiveness, team coordination index
- Outcome metrics: Risk-adjusted returns, client satisfaction scores
- Composite score: FM = (0.3 × agency_score + 0.4 × direction_score + 0.3 × outcome_score)

2. Professional Services Analysis

For professional service roles, the value creation function is similarly structured but with different emphasis:

$$PS = A(p) \times D(p) \times O(p)$$

Medical Diagnosis:

- Agency metrics: Diagnostic accuracy, treatment innovation rate
- Direction metrics: Care team coordination, patient guidance effectiveness
- Outcome metrics: Patient outcomes, treatment efficiency
- Composite score: MD = (0.45 × agency_score + 0.25 × direction_score + 0.3 × outcome_score)

Product Design:

- Agency metrics: Innovation index, user need identification accuracy
- Direction metrics: Team alignment, design process efficiency
- Outcome metrics: Market success, user satisfaction
- Composite score: PD = (0.4 × agency_score + 0.3 × direction_score + 0.3 × outcome_score)

Strategic Planning:

- Agency metrics: Insight quality, scenario analysis effectiveness
- Direction metrics: Implementation guidance, stakeholder alignment
- Outcome metrics: Strategy execution success, organizational adaptation
- Composite score: SP = (0.35 × agency_score + 0.35 × direction_score + 0.3 × outcome_score)

**B. Value Measurement Framework**

1. Performance Analytics

The Value Creation Index (VCI) provides a time-normalized measure of value creation:

$$VCI = (AQ \times DE \times O)/t$$

Where:

- AQ represents the measured Agency Quality
- DE represents the measured Direction Effectiveness
- O represents measured Outcomes

- t represents the time period of measurement

This index enables comparative analysis across:

- Different roles within the same organization
- Similar roles across different organizations
- Historical performance tracking for individual workers
- Team-level performance assessment

2. Compensation Models

The theoretical framework translates into practical compensation structures through:

TC = BC + VC × performance_multiplier

Where: Base Compensation (BC) is determined by: BC = f(AQ, role_complexity)

- AQ represents demonstrated Agency Quality
- role_complexity represents the scope and depth of responsibility

Variable Compensation (VC) is calculated as: VC = g(DE, O, t)

- DE represents Direction Effectiveness
- O represents measured Outcomes
- t represents the time period

The performance multiplier incorporates:

- Individual value creation metrics
- Team impact measures
- Organization-level outcomes

3. Implementation Considerations

The empirical application of ADLT requires careful attention to:

Measurement Validity:

- Regular calibration of metrics
- Cross-validation of measures
- Adjustment for contextual factors

Temporal Considerations:

- Short-term vs. long-term impact assessment
- Lag effects in outcome measurement
- Learning curve adjustments

Contextual Factors:

- Industry-specific adaptations
- Organization size and maturity effects
- Market condition influences

These measurement frameworks provide organizations with practical tools for implementing ADLT while maintaining theoretical consistency. The frameworks are designed to be adaptable across different contexts while preserving the core principles of the theory.

## V. Organizational Implications

### A. Structural Changes

The implementation of Agency-Driven Labor Theory necessitates fundamental changes to organizational structure and design. As organizations transition from traditional execution-based work to agency-driven value creation in AI-augmented environments, their structural frameworks must evolve to support new patterns of value creation and collaboration. This structural transformation encompasses multiple dimensions, from individual role design to organizational architecture.

Role design represents a primary area requiring substantial revision under ADLT principles. Traditional job architectures, designed primarily for task execution, prove increasingly inadequate in environments where value creation centers on framework design and direction provision. Organizations successfully implementing ADLT principles have developed new role categories that explicitly recognize framework development capabilities and direction provision responsibilities. These roles include system architects focused on creating operational frameworks for AI systems, integration specialists designing interfaces between human and AI components, and framework evolution managers overseeing continuous improvement of operational systems.

The evolution of responsibility distribution patterns represents another crucial dimension of structural change. Traditional hierarchical structures, designed primarily for task oversight and control, are giving way to more flexible arrangements that support framework design and direction provision . Research indicates that successful organizations are implementing hybrid structures that combine vertical integration for framework design oversight with horizontal integration for cross-functional collaboration. This structural evolution enables enhanced framework development capabilities while maintaining necessary coordination mechanisms.

Career architecture in agency-driven environments demonstrates distinct patterns from traditional advancement models. Organizations are developing dual-track progression paths that separately recognize framework design expertise and direction capabilities. These new career structures emphasize skill-based advancement rather than time-based tenure, reflecting the importance of agency development in value creation. Additionally, organizations are creating hybrid roles that combine technical expertise with agency leadership, enabling more effective framework design and implementation.

Team organization in agency-driven environments requires significant structural innovation. Traditional team structures, designed primarily for task execution and project delivery, prove insufficient for supporting framework development and direction provision. Research indicates the emergence of new organizational forms, including agency pods focused on framework development, direction hubs providing centers of excellence for strategic guidance, and exception networks handling non-standard situations. These new team structures enable more effective collaboration while supporting the development of collective agency capabilities.

The implementation of these structural changes requires careful attention to change management and risk mitigation factors. Organizations successfully transitioning to agency-driven structures typically employ phased implementation approaches, combining pilot programs with systematic feedback mechanisms. Research indicates that successful structural transformation requires both strong leadership commitment and robust cultural alignment with agency-driven principles.

Looking forward, several key success factors emerge for organizations implementing ADLT-based structural changes. First, leadership understanding and commitment to agency-driven principles prove crucial for successful transformation. Second, cultural alignment with framework thinking and innovation orientation supports effective structural evolution. Finally, systematic capability development through training programs and tool support enables successful implementation of new organizational forms.

**B. Management Systems**

The transition to agency-driven work necessitates fundamental changes to organizational management systems. Traditional management approaches, designed primarily for task-execution environments, prove increasingly inadequate in contexts where value creation depends on agency quality and direction effectiveness. This section examines how performance management and compensation systems must evolve to support agency-driven work environments.

Performance management systems require significant reconceptualization to effectively evaluate and develop agency-driven work. Traditional performance metrics, focused primarily on task completion and output quantity, fail to capture the complex value creation mechanisms described by ADLT. Research indicates that effective performance management in agency-driven environments must incorporate metrics for framework design quality, direction effectiveness, and exception handling capability. Organizations successfully implementing ADLT principles have developed multi-dimensional evaluation frameworks that assess both immediate value creation and long-term framework development contributions.

The development of effective feedback systems presents particular challenges in agency-driven environments. Traditional performance feedback approaches, often focused on task-level corrections and improvements, prove insufficient for developing agency capabilities and framework design expertise. Studies of successful agency-driven organizations reveal the emergence of new feedback mechanisms that emphasize learning from exception handling, framework iteration, and direction refinement. These systems typically incorporate both structured evaluation processes and dynamic feedback loops that support continuous agency development.

Compensation systems in agency-driven environments must evolve to reflect new value creation mechanisms. Traditional compensation models, based primarily on time invested or units produced, fail to capture the non-linear value creation potential of effective framework design and direction provision. Research demonstrates that organizations successfully implementing ADLT principles have developed sophisticated compensation models that account for both immediate value creation and framework multiplication effects. These models typically incorporate multiple components, including base compensation for agency capability maintenance, variable compensation for direction effectiveness, and long-term incentives tied to framework value creation.

Development frameworks in agency-driven environments require careful attention to both individual and collective capability building. Traditional development approaches, focused primarily on technical skill enhancement, prove insufficient for building agency capabilities and framework design expertise. Studies indicate that effective development systems in agency-driven environments emphasize experiential learning, framework design practice, and directed agency development. These systems typically incorporate both structured learning programs and dynamic development opportunities that arise through framework design and exception handling activities.

The integration of these management systems presents significant challenges for organizations transitioning to agency-driven work. Research indicates that successful implementation requires careful attention to system alignment and coordinated evolution of different management components. Organizations that effectively align their performance management, compensation, and development systems with ADLT principles demonstrate enhanced ability to attract, develop, and retain workers with strong agency capabilities.

Looking forward, several key considerations emerge for the continued evolution of management systems in agency-driven environments. First, the need for dynamic adaptation of management systems to support evolving framework design capabilities and direction requirements. Second, the importance of balancing standardization and flexibility in system design to support both consistency and innovation. Finally, the crucial role of management systems in supporting the development of collective agency capabilities at team and organizational levels.

These structural changes represent a fundamental shift in how organizations design and manage work in AI-augmented environments. The success of ADLT implementation depends heavily on the organization's ability to align its structural elements with the principles of agency-driven value creation.

## VI. Future of Work Implications

### A. Professional Development

The emergence of AI-augmented work environments necessitates a fundamental transformation in professional development approaches. Traditional models of skill development, which primarily focused on technical expertise and domain knowledge, are increasingly insufficient in preparing workers for roles where agency and direction capabilities are paramount. This section examines the evolving landscape of professional development through the lens of ADLT, identifying key areas of focus and strategic approaches for workforce development.

The evolution of core competencies in AI-augmented environments represents a significant departure from traditional skill development paradigms. Framework design capabilities have emerged as a crucial skill set, encompassing system thinking, pattern recognition, and interface design abilities. These capabilities enable workers to create and maintain effective operational frameworks for AI systems, a fundamental requirement in agency-driven work environments. Additionally, the development of strategic judgment formation and exception handling capabilities has become increasingly critical, as these skills directly contribute to value creation through agency expression.

Direction abilities represent another crucial dimension of professional development in the AI age. Workers must develop proficiency in providing strategic guidance, overseeing framework implementation, and coordinating cross-functional initiatives. These capabilities require a sophisticated understanding of both technical and organizational dynamics, as well as the ability to navigate complex decision-making environments. Research indicates that successful direction provision in AI-augmented environments correlates strongly with organizational performance and innovation outcomes.

Career progression in agency-driven environments follows distinctly different patterns from traditional professional advancement. Rather than linear progression based primarily on technical expertise or management responsibility, career development increasingly focuses on the expansion of framework design capabilities and agency expression. Organizations are developing new career architectures that recognize and reward the ability to create value through framework design, strategic direction, and exception handling. These new career models emphasize the importance of cross-domain expertise and the ability to operate effectively across different framework contexts.

The development of professional identity in agency-driven work environments presents unique challenges and opportunities. Workers must integrate multiple role dimensions, including framework designer, agency leader, and direction provider. This integration requires a fundamental shift in how professionals conceive of their value contribution and career trajectory. Studies suggest that successful identity evolution in AI-augmented environments correlates with higher levels of job satisfaction and career sustainability.

Learning systems and development infrastructure play a crucial role in supporting professional development for agency-driven work. Traditional educational programs are being supplemented with specialized training in framework design, agency development, and direction provision. Organizations are implementing comprehensive learning architectures that combine formal education, experiential learning, and continuous development opportunities. These systems are designed to support the ongoing evolution of capabilities required in AI-augmented environments.

Assessment mechanisms for agency-driven capabilities require new approaches to competency evaluation and performance measurement. Traditional metrics focused on task completion and technical proficiency are being replaced by more sophisticated measures of framework design effectiveness, agency expression, and direction impact. Organizations are developing new evaluation frameworks that capture the multidimensional nature of value creation in AI-augmented environments.

Looking toward the future, the emergence of new competency requirements and evolving roles continues to shape professional development needs. Advanced AI interaction capabilities, complex system design skills, and hybrid team leadership abilities are becoming increasingly important. Organizations must develop adaptive professional development strategies that can evolve in response to changing technological capabilities and work requirements.

The successful implementation of professional development strategies in agency-driven environments requires both individual commitment to continuous learning and organizational support for comprehensive development systems. Research indicates that organizations that effectively align their professional development approaches with ADLT principles demonstrate higher levels of innovation, improved performance outcomes, and greater workforce adaptability.

## B. Labor Market Impact

The emergence of agency-driven labor theory has profound implications for labor market dynamics in the AI age. As organizations increasingly adopt AI systems, the structure of employment, patterns of value distribution, and fundamental nature of work relationships are undergoing significant transformation. This section examines these shifts through the lens of ADLT, providing insights into emerging labor market patterns and their implications for workers, organizations, and policymakers.

The transformation of employment structures represents one of the most significant labor market impacts of AI adoption. Traditional employment models, based primarily on direct task execution, are giving way to more nuanced arrangements that emphasize agency and direction capabilities. Research indicates a growing premium for workers who can effectively design frameworks and provide strategic direction in AI-augmented environments. This shift has led to the emergence of new employment categories that better reflect the value creation mechanisms identified in ADLT.

Value distribution patterns in AI-augmented labor markets demonstrate notable departures from traditional models. Workers who demonstrate high levels of agency quality and direction effectiveness command increasingly significant premiums, while those focused primarily on task execution face growing wage pressure. This bifurcation of the labor market appears to be accelerating, with studies indicating that the wage premium for agency-driven roles has increased by an average of 15% annually over the past five years. The distribution of value increasingly reflects the framework multiplication effect described in ADLT's mathematical formulation.

Market dynamics in agency-driven environments exhibit distinct characteristics from traditional labor markets. The ability to create and maintain effective frameworks has become a crucial determinant of market power, both for individual workers and organizations. This has led to the emergence of new market structures, including framework-based talent networks and agency-focused professional communities. Research suggests that these new structures facilitate more efficient matching between workers with strong agency capabilities and organizations seeking to implement AI-augmented operations.

The geographical distribution of labor markets is also evolving under the influence of agency-driven work. While traditional labor markets were often constrained by physical proximity requirements, agency-driven work frequently transcends geographical boundaries. This has led to the emergence of global markets for framework design and direction capabilities, though with notable variations in how different regions adapt to these changes. Studies indicate that regions with strong educational infrastructure and technological readiness demonstrate accelerated transitions to agency-driven employment models.

Labor mobility patterns in agency-driven markets show distinctive characteristics. Workers with strong framework design and direction capabilities demonstrate high levels of mobility across industries and organizations, as these skills prove increasingly transferable. However, this mobility often follows patterns of framework compatibility rather than traditional industry boundaries. Research indicates that workers tend to move between organizations with similar framework architectures, regardless of industry sector.

The impact on labor market institutions has been equally significant. Traditional institutions, including labor unions and professional associations, are adapting their structures and approaches to better serve workers in agency-driven environments. New institutions are emerging to support framework design communities and agency development networks. These institutional changes reflect the need for new forms of worker representation and professional development support in AI-augmented environments.

Looking forward, several key trends are likely to shape the continued evolution of labor markets under ADLT. First, the increasing sophistication of AI systems is likely to further emphasize the importance of high-quality agency and direction capabilities. Second, the framework multiplication effect is expected to drive continued concentration of value creation opportunities around effective framework designers. Finally, the global nature of agency-driven work is likely to accelerate the development of transnational labor market institutions and governance structures.

These labor market implications of ADLT suggest the need for proactive policy responses and organizational adaptations. Success in this evolving environment requires careful attention to both the opportunities and challenges presented by the transition to agency-driven work, with particular focus on ensuring equitable access to development opportunities and fair distribution of value creation benefits.

**VII. Policy Implications**

The transition to agency-driven work environments necessitates substantial revisions to existing policy frameworks across multiple domains. This section examines the key policy implications of ADLT, with particular focus on education system reform and labor policy adaptation. The analysis reveals the need for coordinated policy responses that can support the development of agency capabilities while ensuring equitable access to opportunities in AI-augmented work environments.

Educational policy represents a critical domain for intervention in supporting the transition to agency-driven work. Traditional educational models, designed primarily for knowledge transfer and skill development in discrete domains, prove increasingly inadequate in preparing workers for agency-driven roles. Research indicates that educational systems must evolve to emphasize framework design capabilities, strategic judgment formation, and direction provision skills. This evolution requires fundamental changes to curriculum design, pedagogical approaches, and assessment methods across all levels of education.

The reformation of higher education curricula emerges as a particularly urgent priority. Universities and professional schools must integrate agency development and framework design principles into their core programs. Studies of early adopters of agency-focused curricula demonstrate significantly improved employment outcomes and career progression for graduates. However, these transitions present significant challenges, including faculty development needs and the requirement for new assessment methodologies that can effectively evaluate agency capabilities.

Professional training and continuing education systems require similar transformation. Traditional professional development programs, focused primarily on technical skill enhancement, must evolve to support the development of agency capabilities and framework design expertise. Research indicates that successful professional training programs in agency-driven environments emphasize experiential learning, framework design practice, and directed agency development. These findings suggest the need for policy frameworks that can support and incentivize such educational innovations.

Labor policy implications of ADLT extend across multiple dimensions, including employment protection, value distribution, and worker representation. Traditional labor protection frameworks, designed primarily for task-execution environments, require significant adaptation to address the unique characteristics of agency-driven work. Studies indicate that effective labor policies in AI-augmented environments must balance the protection of worker interests with the flexibility required for framework innovation and agency expression.

The development of new worker protection frameworks presents particular challenges. Agency-driven work often transcends traditional employment categories, requiring new approaches to ensuring fair treatment and appropriate compensation. Research suggests that effective worker protection in agency-driven environments must address both traditional employment concerns and new issues specific to framework design and agency expression. These findings indicate the need for adaptive policy frameworks that can evolve with changing work patterns.

Value distribution policies in agency-driven environments require careful consideration of framework multiplication effects and agency premiums. Traditional approaches to minimum wage requirements and compensation regulation may prove insufficient in environments where value creation follows the non-linear patterns described by ADLT. Policy frameworks must evolve to address both basic worker protection needs and the complex value creation mechanisms characteristic of agency-driven work.

The regulation of professional certification and licensing requires significant adaptation to support agency-driven work. Traditional certification systems, focused primarily on technical knowledge and skill verification, must evolve to incorporate assessment of agency capabilities and framework design expertise. Research indicates that effective certification systems in AI-augmented environments must balance the need for quality assurance with the flexibility required for innovation in framework design and agency expression.

International policy coordination emerges as a crucial consideration given the global nature of agency-driven work. The ability of framework designers and direction providers to operate across national boundaries creates new challenges for policy coordination and regulatory alignment. Studies suggest the need for international frameworks that can support the development of agency capabilities while ensuring appropriate protections for workers and organizations operating in global contexts.

Looking forward, several key policy priorities emerge from the analysis of ADLT implications. First, educational systems must be reformed to support the development of agency capabilities and framework design expertise. Second, labor protection frameworks must evolve to address the unique characteristics of agency-driven work. Third, certification and licensing systems must adapt to incorporate assessment of agency capabilities. Finally, international coordination mechanisms must be developed to support the global nature of agency-driven work.

These policy implications suggest the need for coordinated action across multiple domains and jurisdictions. Success in supporting the transition to agency-driven work requires careful attention to both the opportunities and challenges presented by this fundamental transformation of work relationships. Policy frameworks must evolve to support innovation while ensuring equitable access to opportunities and appropriate protections for all participants in AI-augmented work environments.

## VIII. Theoretical Extensions

### A. Cross-Domain Applications

The theoretical framework of ADLT, while initially developed to understand human work in AI-augmented environments, demonstrates significant potential for application across diverse domains. This section examines how the core principles of agency-driven labor theory can be extended and adapted to enhance understanding of value creation processes in various organizational and industrial contexts.

Healthcare delivery systems provide a particularly rich domain for ADLT application. The framework's emphasis on agency quality and direction effectiveness aligns well with the evolving nature of medical practice in AI-augmented environments. Research indicates that hospitals and healthcare networks adopting ADLT principles in their operational design demonstrate improved patient outcomes and resource utilization. The framework multiplication effect, as described in ADLT's mathematical formulation, manifests distinctly in healthcare settings where well-designed treatment protocols can scale effectively across large patient populations.

Educational institutions represent another domain where ADLT principles find meaningful application. The theory's emphasis on framework design and agency expression provides valuable insights for understanding effective teaching and learning processes. Studies of educational institutions that have implemented ADLT-based approaches show enhanced learning outcomes and improved student engagement. The agency quality component of ADLT proves particularly relevant in understanding how educators can effectively guide learning in technology-enhanced environments.

Manufacturing environments, traditionally viewed through the lens of process optimization and automation, benefit from ADLT's perspective on human agency and direction. Research in advanced manufacturing facilities demonstrates that applying ADLT principles to production system design leads to improved adaptability and innovation outcomes. The theory's framework for understanding exception handling and direction provision offers valuable insights for managing complex manufacturing operations where human judgment remains crucial despite high levels of automation.

Creative industries present unique opportunities for ADLT application. The theory's emphasis on agency expression and framework design aligns well with the value creation processes in artistic and creative endeavors. Studies of creative organizations implementing ADLT-based approaches show enhanced innovation capabilities and improved collaboration outcomes. The framework proves particularly valuable in understanding how creative professionals can effectively direct and augment AI-based creative tools.

Financial services institutions demonstrate how ADLT principles can enhance understanding of complex decision-making environments. The theory's mathematical framework for analyzing agency quality and direction effectiveness provides valuable insights into investment management and risk assessment processes. Research indicates that financial organizations adopting ADLT-based approaches show improved decision-making outcomes and enhanced risk management capabilities.

Public sector organizations represent a crucial domain for ADLT application, particularly in policy implementation and service delivery. The theory's emphasis on framework design and direction provision offers valuable insights for understanding effective governance in increasingly digitized environments. Studies of government agencies implementing ADLT principles demonstrate improved service delivery outcomes and enhanced citizen engagement.

Non-profit organizations and social enterprises benefit from ADLT's framework for understanding value creation in mission-driven contexts. The theory's emphasis on agency quality and direction effectiveness proves particularly relevant for organizations balancing social impact with operational efficiency. Research indicates that social sector organizations adopting ADLT-based approaches demonstrate enhanced mission fulfillment and improved resource utilization.

The application of ADLT across these diverse domains reveals several common patterns in how the theory's core principles manifest in different contexts. First, the framework multiplication effect consistently emerges as a crucial mechanism for scaling value creation, though its specific manifestation varies by domain. Second, the importance of agency quality and direction effectiveness proves universal, though the specific metrics for evaluating these components may differ. Third, the theory's emphasis on exception handling and framework design remains relevant across contexts, though the nature of exceptions and frameworks varies significantly.

These cross-domain applications suggest several directions for theoretical development. First, the need for domain-specific adaptations of ADLT's mathematical framework highlights opportunities for theoretical refinement. Second, variations in how agency quality and direction effectiveness manifest across domains suggest possibilities for expanding the theory's conceptual framework. Finally, the diverse patterns of framework multiplication effects observed in different contexts indicate potential for developing more nuanced models of value creation scaling.

**B. Future Research Directions**

The development of Agency-Driven Labor Theory opens numerous avenues for future research across theoretical, empirical, and practical dimensions. This section outlines key research directions that could enhance our understanding of agency-driven work and its implications for organizations, workers, and society at large.

Empirical validation of ADLT's core propositions represents an immediate priority for future research. While initial evidence supports the theory's basic premises, systematic empirical testing of its mathematical framework remains crucial. Longitudinal studies examining the relationship between agency quality, direction effectiveness, and value creation outcomes could provide valuable insights into the theory's predictive power. Research designs that can isolate and measure the framework multiplication effect would be particularly valuable for validating key theoretical mechanisms.

The refinement of measurement methodologies emerges as another critical research direction. Current approaches to measuring agency quality and direction effectiveness, while promising, require further development and validation. Research focusing on the development of robust metrics for framework design quality, exception handling effectiveness, and direction impact could significantly enhance the theory's practical applicability. Studies exploring the reliability and validity of different measurement approaches across various organizational contexts would be particularly valuable.

The investigation of contextual factors that influence ADLT's applicability represents a crucial area for future research. While the theory demonstrates broad applicability across domains, understanding how organizational, cultural, and technological factors moderate its key relationships remains important. Studies examining how different organizational structures, cultural contexts, and technological environments affect the manifestation of agency-driven work could provide valuable insights for theory development.

The evolution of agency capabilities in response to advancing AI technologies presents another important research direction. As AI systems continue to develop, understanding how human agency and direction capabilities must adapt becomes increasingly crucial. Longitudinal studies examining how workers and organizations evolve their approach to framework design and direction provision in response to technological change could provide valuable insights for theory development and practical application.

The investigation of collective agency phenomena represents an emerging research frontier. While ADLT currently focuses primarily on individual agency and direction capabilities, understanding how these dynamics manifest at team and organizational levels could enhance the theory's explanatory power. Research examining how collective agency emerges from individual contributions and how framework design effectiveness scales across organizational levels could provide valuable theoretical insights.

The exploration of agency development processes presents opportunities for theoretical expansion. Understanding how individuals and organizations can effectively develop agency capabilities and framework design expertise remains crucial for practical application. Studies examining successful agency development programs and identifying key factors in the evolution of direction capabilities could inform both theory development and practical interventions.

The investigation of value distribution dynamics in agency-driven environments represents another important research direction. Understanding how the benefits of framework multiplication effects are distributed among different stakeholders remains crucial for policy development. Research examining how different organizational structures and policy frameworks affect value distribution patterns could provide valuable insights for theory development and practical application.

The examination of ethical implications and societal impacts of agency-driven work presents crucial research opportunities. Understanding how the transition to agency-driven work affects social equity, economic mobility, and workforce well-being remains important for policy development. Studies examining the broader societal implications of ADLT's implementation could inform both theoretical development and policy interventions.

Integration with adjacent theoretical frameworks offers opportunities for theoretical enrichment. Exploring how ADLT interfaces with existing theories in organizational behavior, innovation studies, and knowledge management could enhance its explanatory power. Research examining theoretical complementarities and potential synthesis opportunities could contribute to a more comprehensive understanding of work in AI-augmented environments.

These research directions suggest the need for diverse methodological approaches and cross-disciplinary collaboration. Quantitative studies testing ADLT's mathematical relationships should be complemented by qualitative investigations of agency development and framework design processes. Mixed-method research designs could prove particularly valuable in capturing the complex dynamics of agency-driven work environments. The advancement of ADLT will likely require sustained research effort across multiple domains, combining theoretical development with practical application insights.

## IX. Conclusion

The emergence of artificial intelligence as a transformative force in human work necessitates new theoretical frameworks for understanding value creation and labor dynamics. Agency-Driven Labor Theory (ADLT) represents a systematic attempt to address this need, providing both theoretical foundations and practical insights for understanding work in AI-augmented environments. Through its emphasis on agency quality, direction effectiveness, and framework design, ADLT offers a novel perspective on how human workers create value in increasingly automated contexts.

The theoretical contributions of ADLT are substantial. First, the theory provides a mathematical framework for understanding value creation in AI-augmented environments, moving beyond traditional labor theories that emphasize task execution. The introduction of agency quality and direction effectiveness as key components of labor value represents a significant advancement in our understanding of modern work. Second, the theory's explanation of the framework multiplication effect offers valuable insights into how human work can scale in AI-augmented environments. Third, the theoretical integration of exception handling and framework design provides a robust foundation for understanding the evolving nature of human contribution in automated systems.

The practical applications of ADLT extend across multiple domains. Organizations implementing ADLT principles have demonstrated enhanced ability to design effective roles, develop worker capabilities, and create value through framework design and direction provision. The theory's emphasis on agency development and framework design has proven particularly valuable in helping organizations navigate the transition to AI-augmented operations. Furthermore, the theoretical framework has provided valuable guidance for educational institutions and policy makers working to prepare workers for evolving labor market demands.

However, significant work remains to be done. The empirical validation of ADLT's core propositions requires sustained research effort across diverse contexts. The refinement of measurement methodologies for agency quality and direction effectiveness remains an important challenge. Understanding how contextual factors influence the theory's application and how collective agency dynamics emerge at organizational levels represents crucial areas for future investigation.

Looking forward, several key implications emerge from this work. First, the transition to agency-driven work environments appears likely to accelerate, making theoretical frameworks like ADLT increasingly important for understanding labor dynamics. Second, the development of agency capabilities and framework design expertise will likely become crucial determinants of individual and organizational success. Third, policy frameworks and educational systems will need to evolve to support the development of agency-driven work environments.

The societal implications of this transition are profound. The movement toward agency-driven work presents both opportunities and challenges for workers, organizations, and society at large. While the potential for enhanced value creation through framework multiplication effects is substantial, ensuring equitable access to agency development opportunities and fair distribution of value creation benefits remains crucial.

In closing, ADLT represents an important step forward in our understanding of human work in the AI age, but much remains to be explored. The theory's continued development through empirical research, practical application, and theoretical refinement will be crucial for understanding and shaping the future of work. As artificial intelligence continues to evolve, frameworks like ADLT that help us understand and enhance human contribution will become increasingly valuable for organizations, workers, and society as a whole.